\documentclass[twocolumn,showpacs,preprintnumbers,amsmath,amssymb,aps]{revtex4}
\usepackage{graphicx}
\usepackage{dcolumn}
\usepackage{bm}
\usepackage{color}
\begin{document}
\preprint{}
\title{Constraints on nuclear matter parameters of an Effective Chiral Model}
\author{T. K. Jha$^{1}${\footnote {email: tkjha@prl.res.in}} and H. Mishra$^{1,2,3}${\footnote {email: hm@prl.res.in}}}
\affiliation {
$1$ Theoretical Physics Division, Physical Research Laboratory, Navrangpura, 
Ahmedabad, India - 380 009 \\
$2$ Institut f\"ur Theoretische Physik, J. W. Goethe Universit\"at, Max-von-Laue-Str. 1, 60438 Frankfurt am Main, Germany \\
$3$ School of Physical Sciences, Jawaharlal Nehru University, New Delhi, India 110067
}
\date{\today}
\begin{abstract}

Within an effective non-linear chiral model, we evaluate nuclear matter parameters exploiting the uncertainties in the nuclear saturation properties. The model is sternly constrained with minimal free parameters, which display the interlink between nuclear incompressibility ($K$), the nucleon effective mass ($m^{\star}$), the pion decay constant ($f_{\pi}$) and the $\sigma-$meson mass ($m_{\sigma}$). The best fit among the various parameter set is then extracted and employed to study the resulting Equation of state (EOS). Further, we also discuss the consequences of imposing constraints on nuclear EOS from Heavy-Ion collision and other phenomenological model predictions.

\end{abstract}
\maketitle
\section{Introduction}

The framework of Quantum Hadrodynamics \cite{wal74,serot86} as an elegant and consistent theoretical treatment of finite nuclei as well as infinite nuclear matter laid down the pillars of relativistic theories which seem to provide solution to the so called ``{\it the Coester band}" problem \cite{co01,ma89}. However, our present knowledge of nuclear matter is confined around nuclear saturation density ($\rho_0 \approx 3 \times 10^{14} g cm^{-3}$) and therefore, in order to have some meaningful correlations while extrapolating to higher densities, the nuclear equation of state (EOS) must satisfy certain minimum criteria quantified as the {\it ``nuclear saturation properties''}, which are the physical constants of nature. Basically it is understood that the inherited uncertainty at $\rho_0$ gets more pronounced at higher densities ($3-10~\rho_0$), relevant to astrophysical context such as the modeling of neutron stars. In this context, the two most important quantities which play vital role and are known to have substantial impact on the EOS are the nucleon effective mass and the nuclear incompressibility \cite{compact,teukolsky}.  Ironically, these two properties are not very well determined and they posses large uncertainty. The nuclear incompressibility derived from nuclear measurements and astrophysical observations exhibit a broad range of values $K = (180 - 800)$ MeV \cite{glen88}. Further the non-relativistic and the relativistic models fails to agree to a commom consensus. The non-relativistic calculations predict the compression moduli in the range $K=(210-240) MeV$ \cite{k2,k3,k4}, whereas, relativistic calculations predicts it in the range $(200-300)~MeV$ \cite{nl3,k6}. Apart from that we are inevitably marred by the uncertainty in the determination of mass of the scalar meson ($\sigma$-meson). The attractive force resulting from the scalar sector is responsible for the intermediate range attraction which, along with the repulsive vector forces provides the saturation mechanism for nuclear matter \cite{serot86}. The estimate from the Particle Data Group quotes the mass of this scalar meson `$f_0 (600)$' or $\sigma-$meson in the range $(400 - 1200)$ MeV \cite{pdg}. A recent estimate however, for sigma meson mass   is found to be $513 \pm 32$ MeV \cite{mura02}. 

Phenomenologically, parallel to the well known $\sigma - \omega$ model, preferably known as the Walecka model \cite{wal74,serot86,boguta77}, chiral models \cite{ch01,ch02,ch03,ch03a,ch04,ch05,ch06} have been developed and were applied to nuclear matter studies. Chiral symmetry is a symmetry of strong interactions in the limit of vanishing quark masses and is desirable in any relativistic theory. However, because the current quark masses are small but finite this symmetry can be considered as an approximate symmetry. This symmetry is spontaneously broken in the ground state. In the context of $\sigma-$models, the $\sigma-$ field (which carries the quantum numbers of the vacuum) attains a finite vacuum expectation value $\langle\sigma\rangle = \sigma_0 = f_{\pi}$. Equivalently, the potential for the $\sigma-$ field attains a minimum at $f_\pi$  \cite{koch,one}. The value of $f_{\pi}$ reflects the strength of the symmetry breaking and experimentally it is found to be $f_{\pi} \approx 131 $ MeV \cite{pdg}.

Time and again, the aforesaid facts and figures emphasize the need to address the importance of imposing constraints to the EOS to narrow down the uncertainties both experimentally and theoretically. Arguably, to address these issues, one needs a model that has the desired attributes of the relativistic framework and which can be successfully applied to various nuclear force problem both in the vicinity of $\rho_0$ as well as at higher densities with the same set of parameters. With this motivation, we choose a model \cite{tkj06} which embodies chiral symmetry and has minimum number of free parameters (total five) to reproduce the saturation properties. The spontaneous breaking of chiral symmetry relates the mass of the hadrons to the vacuum expectation value of the scalar field and thus naturally restricts the parameters of the model. Therefore, the present study, apart from testing the reliability of the model, puts valuable constraint on the EOS based on the pion decay constant and brings out correlations between between the pion decay constant ($f_{\pi}$), the $\sigma-$meson mass ($m_{\sigma}$), the nuclear incompressibility ($K$) and the nucleon effective mass ($m^{\star}$).

In section 2, we briefly describe basic ingredients of the hadronic model and the energy and pressure of many baryonic system is computed following the mean-field ansatz. Subsequent section (Section 3) describes the methodology to evaluate the model parameters. In Section 4, we extract the best fit among the various parameter of the model and apply it to study the resulting EOS of symmetric nuclear matter. In the result and discussion section, we discuss the consequences of imposing various constraints on the model parameters and finally, we conclude with some important findings of this work.

\section{The Effective Chiral Model}

Using the chiral sigma model with dynamically generated mass for vector meson, Glendening studied finite temperature aspects of nuclear matter and its application to neutron stars \cite{gl}. However, there the $\rho-$meson and its isospin symmetry influence was not considered. Although a nice framework respecting chiral symmetry, a drawback was its unacceptable high incompressibility and in the subsequent extension of the model \cite{five}, the mass of the vector meson is not generated dynamically. The model that we consider \cite{tkj06} in our present analysis embodies higher orders of the scalar field in addition to the dynamically generated mass of the vector meson. Without higher order in scalar field interactions, the model was first employed to study high density matter \cite{dutta}. To bring down the resulting high incompressibility, non-linear interaction in the scalar field was included in later work \cite{sahu} and subsequently applied to study nuclear matter at finite temperature \cite{tkj04}. The success of the model then motivated us to generalize it to include the octet of baryons and to study hyperon rich matter and properties of neutron star \cite{tkj06,tkj08}. However, in earlier works, the parameter sets that were employed were not studied and analyzed in detail with respect to the inherent vacuum properties of chiral symmetry. Moreover, rather than a phenomenological fit the parameters must be constrained meaningfully, so that the resulting EOS is more realistic and purposeful. Motivated by this, we presently try to explore the consequences of imposing stringent constraint on the model parameters not only with properties known at saturation density but also on the resulting EOS with other phenomenological model predictions and experimental data at high density. In addition to that, the correlation between various quantities with the vacuum value of the scalar field naturally spells out definite interlink between them. We now proceed to describe the salient features of the present model. The effective Lagrangian of the model interacting through the exchange of the pseudo-scalar meson $\pi$, the scalar meson $\sigma$, the vector meson $\omega$ and the iso-vector $\rho-$meson is given by:

\begin{widetext}
\begin{eqnarray}
\label{lag}
{\cal L}&=& \bar\psi_{B}~\left[ \big(i\gamma_\mu\partial^\mu
         - g_{\omega}\gamma_\mu\omega^\mu
         - \frac{1}{2}g_{\rho}{\vec \rho}_\mu\cdot{\vec \tau}
            \gamma^\mu\big )
         - g_{\sigma~}~\big(\sigma + i\gamma_5
             \vec \tau\cdot\vec \pi \big)\right]~ \psi_{B}
\nonumber \\
&&
        + \frac{1}{2}\big(\partial_\mu\vec \pi\cdot\partial^\mu\vec\pi
        + \partial_{\mu} \sigma \partial^{\mu} \sigma\big)
        - \frac{\lambda}{4}\big(x^2 - x^2_0\big)^2
        - \frac{\lambda b}{6 m^2}\big(x^2 - x^2_0\big)^3
        - \frac{\lambda c}{8 m^4}\big(x^2 - x^2_0\big)^4
\nonumber \\
&&      - \frac{1}{4} F_{\mu\nu} F_{\mu\nu}
        + \frac{1}{2}{g_{\omega B}}^{2}x^2 \omega_{\mu}\omega^{\mu}
        - \frac {1}{4}{\vec R}_{\mu\nu}\cdot{\vec R}^{\mu\nu}
        + \frac{1}{2}m^2_{\rho}{\vec \rho}_{\mu}\cdot{\vec \rho}^{\mu}\ .
\end{eqnarray}
\end{widetext}

The first line of the above Lagrangian represents the interaction of the nucleon isospin doublet $\psi_B$ with the aforesaid mesons. In the second line we have the kinetic and the
non-linear terms in the pseudo-scalar-isovector pion field `$\vec \pi$', the scalar field `$\sigma$', and higher order terms of the scalar field in terms of the invariant combination of the two i.e., $x^2= {\vec \pi}^2+\sigma^{2}$. Finally in the last line, we have the field strength and the mass term for the vector field `$\omega$' and the iso-vector field `$\vec \rho$' meson. $g_{\sigma}, g_{\omega}$ and $g_{\rho}$ are the usual meson-nucleon coupling strength of the scalar, vector and the iso-vector fields respectively. Here we shall be concerned only with the normal non-pion condensed state of matter, so we take $<\vec \pi>=0$ and also $m_{\pi} = 0$.

The interaction of the scalar and the pseudoscalar mesons with the vector boson generates a dynamical mass for the vector bosons through spontaneous breaking of the chiral symmetry with scalar field attaining the vacuum expectation value $x_0$. Then the mass of the nucleon ($m$), the scalar ($m_{\sigma}$) and the vector meson mass ($m_{\omega}$), are related to $x_0$ through

\begin{eqnarray}
m = g_{\sigma} x_0,~~ m_{\sigma} = \sqrt{2\lambda} x_0,~~
m_{\omega} = g_{\omega} x_0\ .
\end{eqnarray}
\noindent

To obtain the equation of state, we revert to the mean-field procedure in which, one assumes the mesonic fields to be uniform i.e., without any quantum fluctuations. We recall here that this approach has been extensively used to obtain field-theoretical EoS for high density matter \cite{five}, and gets increasingly valid when the source terms are large \cite{serot86}. The details of the present model and its attributes such as the derivation of the equation of motion of the meson fields and its equation of state $(\varepsilon~\&~P)$ can be found in our preceding work \cite{tkj06,tkj08}. For the sake of completeness however, we write down the meson field equations in the mean-field ansatz. The vector field ($\omega$), the scalar field ($\sigma$) (in terms of $Y=x/x_0 = m^{\star}/m$) and the isovector field ($\rho$) is respectively given by

\begin{eqnarray}
\omega_0=\sum_{B}\frac{ \rho_B }{g_{\omega} x^2},
\end{eqnarray}

\begin{widetext}
\begin{equation}
(1-Y^2) -\frac{b}{m^2 c_{\omega}}(1-Y^2)^2
+\frac{c}{m^4c_{\omega}^2}(1-Y^2)^3
+\frac{2 c_{\sigma}c_{\omega}\rho_B^2}{m^2Y^4}
-\frac{2 c_{\sigma}\rho_S}{m Y}=0\,
\label{effmass}
\end{equation}
\end{widetext}

\begin{equation}
\rho_{03} =\sum_{B} \frac{g_{\rho}}{m_\rho^2} I_{3}~\rho_{B}.
\end{equation}

The quantity $\rho_B$ and $\rho_S$ are the vector and the scalar density defined as,
\begin{equation}
\rho_B= \frac{\gamma}{(2\pi)^3}\int^{k_F}_o d^3k,
\end{equation}

\begin{equation}
\rho_{S}= \frac{\gamma}{(2\pi)^3}\int^{k_F}_o\frac{m^{\star} d^3k}
         {\sqrt {k^2+m^{\star 2}}}.
\end{equation}
\noindent
In the above, `$k_F$' is the fermi momenta of the baryon and $\gamma=4$ (symmetric matter) is the spin degeneracy factor. For symmetric nuclear matter ($N = Z$), we neglect the contribution from the $\rho-$meson. The nucleon effective mass is then $m^{\star} \equiv Y m$ and $c_{\sigma}\equiv  g_{\sigma}^2/m_{\sigma}^2 $ are $c_{\omega} \equiv g_{\omega}^2 /m_{\omega}^2 $ are the scalar and vector coupling parameters that enters in our calculations.

The total energy density `$\varepsilon$' and pressure `$P$' of symmetric nuclear matter for a given baryon density is:

\begin{widetext}
\begin{eqnarray}
\varepsilon
&=&
	  \frac{\gamma}{2\pi^2} \int _o^{k_F} k^2dk\sqrt{{k}^2 + m^{\star 2}}
        + \frac{m^2(1-Y^2)^2}{8c_{\sigma}}
        - \frac{b}{12c_{\omega}c_{\sigma}}(1-Y^2)^3
        + \frac{c}{16m^2c_{\omega}^2c_{\sigma}}(1-Y^2)^4
        + \frac{c_{\omega} \rho_B^2}{2Y^2}
\\
P &=&	\frac{\gamma }{6\pi^2}
                \int _o^{k_F} \frac{k^4dk}{\sqrt{{k}^2 + m^{\star 2}}}
        - \frac{m^2(1-Y^2)^2}{8c_{\sigma}}
        + \frac{b}{12c_{\omega}c_{\sigma}}(1-Y^2)^3
        - \frac{c}{16m^2c_{\omega}^2c_{\sigma}}(1-Y^2)^4
        + \frac{c_{\omega}\rho_B^2}{2Y^2}
\end{eqnarray}
\end{widetext}

The meson field equations for $\omega$ (eqn 3) and $\sigma$-meson (eqn. 4) are solved
self-consistently at a fixed baryon density to obtain the respective field strengths and the corresponding energy density and pressure is calculated.

\section{Evaluation of model parameters}

Having calculated the the thermodynamic quantities such as the energy density and the pressure, our primary aim is to evaluate the set of parameters for the EoS that satisfies the nuclear matter properties defined at normal nuclear matter density ($\rho_0$) at zero temperature. As discussed earlier, a desirable and valid EoS must satisfy the saturation properties of symmetric nuclear matter and the parameters of the model can be adjusted to fit those. Similar procedure has been adopted in Ref. \cite{param1,param2} to evaluate the parameters of the mean-field models.

What we have in our hand is the set of five saturation properties of nuclear matter that a EOS has to satisfy, they are the Binding energy per nucleon ($\approx -16.3$)MeV, the saturation density ($\rho_0 \approx 0.153~ fm^{-3}$), the nuclear incompressibility ($167 - 380$)MeV, the nucleon effective mass ($m^{\star}/m = 0.75 - 0.90$) and the asymmetry energy coefficient ($J = 32 \pm 4$)MeV, all defined at $\rho_0$, the nuclear saturation density. However, it can be seen that the uncertainty in their values enables us to extract and study the parameters within the specified range or with the variation thereof, in order to analyze their effect on a particular EOS. The five parameters of the present model that are to be evaluated are the three meson-nucleon coupling constants ($C_{\sigma}, C_{\omega}, C_{\rho}$) and the two higher order scalar field constants ($b$ \& $c$). 

The individual contributions to the energy density for symmetric nuclear matter (eqn. (8)) can be abbreviated as,

\begin{eqnarray}
\varepsilon=\varepsilon_k+\varepsilon_{\sigma}+\varepsilon_{\omega},
\end{eqnarray}

where,
\begin{eqnarray}
\varepsilon_k=\frac{\gamma}{2\pi^2}\int _o^{k_F} k^2dk\sqrt{{k}^2 
+ m^{\star 2}}\ ,
\end{eqnarray}

\begin{eqnarray}
\varepsilon_{\sigma}&=&\frac{m^2(1-Y^2)^2}{8c_{\sigma}}
        - \frac{b}{12c_{\omega}c_{\sigma}}(1-Y^2)^3
\nonumber \\
        &+& \frac{c}{16m^2c_{\omega}^2c_{\sigma}}(1-Y^2)^4 ,
\end{eqnarray}

and

\begin{eqnarray}
\varepsilon_{\omega}=\frac{c_{\omega} \rho_B^2}{2Y^2},
\end{eqnarray}
\noindent
where $\rho_B = \rho_n + \rho_p$ is the total baryon density which is the sum of the neutron density `$\rho_n$' and the proton density `$\rho_p$'. The relative neutron excess is then given by $\delta = (\rho_n - \rho_p)/\rho_B$. At the standard state $\rho_B = \rho_0$, the nuclear matter saturation density and $\delta = 0$. Consequently, the standard state is then specified by the argument ($\rho_0, 0$), and the energy per particle is $e (\rho_0, 0)$ = $\varepsilon/\rho_0$ - m = $a_1$ = -16.3 MeV for symmetric nuclear matter. The nuclear matter EOS derived earlier can be expressed in terms of the nuclear energy density $\varepsilon$ as,

\begin{eqnarray}
\varepsilon = \varepsilon_k+\varepsilon_{\sigma}+\varepsilon_{\omega}=\rho_0(m-a_1).
\end{eqnarray}

From the the equilibrium condition $P (\rho_0, 0)=0$, we have,

\begin{eqnarray}
P &=& -\varepsilon~+~ \rho_B~\frac{\partial \varepsilon}{\partial \rho_B}~\nonumber \\
&=&~ \frac{1}{3} \varepsilon_k ~-~ \frac{1}{3} m^{\star} \rho_S ~-~ \varepsilon_{\sigma} ~+~ \varepsilon_{\omega} ~=~ 0.
\end{eqnarray}

Consequently, the respective energy contributions can be expressed in terms of
these specified values at the saturation density. Using eqn. (14) and eqn. (15), they are given as, 

\begin{eqnarray}
\varepsilon_{\sigma}=\frac{1}{2} ~\left[\rho_0(m-a_1)-\frac{1}{3}
(2\varepsilon_k+m^{\star}\rho_s)\right]
\end{eqnarray}
\noindent
and

\begin{eqnarray}
\varepsilon_{\omega}=\frac{1}{2}\left[\rho_0(m-a_1)-\frac{1}{3}
(4\varepsilon_k-m^{\star}\rho_s)\right],
\end{eqnarray}

where $\rho_s$ is the scalar density defined in eqn. (7), analytically which is given by,

\begin{eqnarray}
\rho_s=\frac{1}{\pi^2}m^{\star} \left[ k_F E_F 
- ln \Big(\frac{k_F+E_F}{m^{\star}}\Big) m^{\star 2}\right].
\end{eqnarray}

In the above equations, $m^{\star}=Ym$ is the effective nucleon mass and $E_F=\sqrt{k_F^2+m^{\star 2}}$ is the effective energy of the nucleon carying momenta $k_F$.

From eqn. (13), the vector coupling ($C_{\omega}$) can be readily evaluated using the relation

\begin{eqnarray}
C_{\omega}=\frac{2 Y^2}{\rho_0^2} \varepsilon_{\omega},
\end{eqnarray}
\noindent
with $\varepsilon_{\omega}$ given by eqn. (17), for a specified value of $Y=m^{\star}/m$ defined at $\rho_0$. 

Similarly, using the equation of motion for the scalar field (eqn. (4)), the scalar coupling can be calculated using the relation  

\begin{widetext}
\begin{eqnarray}
C_{\sigma} &=& \frac{m Y}{2 \rho_S} \left[(1-Y^2) - \frac{b}{m^2 c_{\omega}}(1-Y^2)^2 + \frac{c}{m^4c_{\omega}^2}(1-Y^2)^3 + \frac{2 c_{\sigma}c_{\omega}\rho_B^2}{m^2Y^4}\right].
\end{eqnarray}
\end{widetext}

In the above expression, the higher order scalar field couplings constants `b' and `c' are unknown, but they can be solved simultaneously to obtain the respective parameters. To compute the constants of the higher order scalar field, we use the equation of motion of the scalar field (eqn. (4)) and eqn. (12). From eqn. (12), we get

\begin{eqnarray}
\frac{c(1-Y^2)}{m^2 c_{\omega}} &=& \frac{16\varepsilon_{\sigma}c_{\omega}
c_{\sigma}}{(1-Y^2)^3} - \frac{2 m^2 c_{\omega}}{(1-Y^2)} +\frac{4 b}{3}.
\end{eqnarray}

From the equation of the motion of scalar field, we get

\begin{eqnarray}
b&=& \frac{2 c_{\sigma} c_{\omega}^2 \rho_B^2}{Y^4 (1-Y^2)^2}
+ \frac{c (1-Y^2)}{m^2 c_{\omega}}
+ \frac{m^2 c_{\omega}}{(1-Y^2)} \nonumber \\
&-& \frac{2 c_{\sigma} c_{\omega} m \rho_S}{Y (1-Y^2)^2}.
\end{eqnarray}

Substituting eqn. (21) in eqn. (22) leads us to the expression to calculate the higher order scalar field constant `$b$', which is,

\begin{eqnarray}
b &=& \frac{6 c_{\sigma} c_{\omega} m \rho_S}{Y (1-Y^2)^2} 
+ \frac{6 c_{\sigma} c_{\omega}^2 \rho_B^2}{Y^4 (1-Y^2)^2}
- \frac{48 \varepsilon_{\sigma} c_{\sigma} c_{\omega}}{(1-Y^2)^3} \nonumber \\
&+& \frac{3 m^3 c_{\omega}}{(1-Y^2)}.
\end{eqnarray}

Similarly, the higher order constant `$c$' in the scalar field can be computed from the relation,

\begin{eqnarray}
c&=& \frac{8 c_{\sigma} c_{\omega}^2 m^3 \rho_s}{Y (1-Y^2)^3}
- \frac{8 c_{\sigma} c_{\omega}^3 m^2 \rho_B^2}{Y^4 (1-Y^2)^3}
- \frac{48 \varepsilon_{\sigma} c_{\sigma} c_{\omega}^2}{(1-Y^2)^4} \nonumber \\
&+& \frac{2 c_{\omega}^2 m^4}{(1-Y^2)^2}.
\end{eqnarray}

The calculation of $C_{\omega}$ is straight forward, but eqn. (20), (23) and (24) can be solved simultaneously numerically for a given initial values of $C_{\sigma}$, b and c, the solution of which would thus return the set of values for a desired value of $Y$ at $\rho_0$.

Finally, for studying asymmetric matter, we need to incorporate the effect of iso-vector $\rho-$meson and the coupling for the $\rho-$ meson has to be obtained by fixing the asymmetry energy coefficient $J \approx 32 \pm 4~ MeV$ \cite{moll88} at $\rho_0$. Accordingly, the $\rho-$ meson coupling constant ($C_{\rho}$) can be fixed using the relation,

\begin{equation}
J = \frac{c_{\rho} k_F^3}{12\pi^2} + \frac{k_F^2}{6\sqrt{(k_F^2 + m^{\star 2})}}\ ,
\end{equation}
where $c_{\rho} \equiv g^2_\rho/m^2_{\rho}$ and $k_F=(6\pi^2\rho_B/\gamma)
^{1/3}$.

Thus the model parameters are evaluated solving equations (19), (20), (23), (24) and (25) self-consistently, for the specified or desired values of the properties of symmetric nuclear matter at saturation point. Further it is also required that the EOS so obtained has a reasonable nuclear incompressibility which is defined as the curvature of the energy curve at the saturation point and is given as,

\begin{equation}
K = 9~\rho_0^2~\frac{\partial^2 (\varepsilon/\rho_B)}{\partial \rho_B} \Big|_0.
\end{equation}

Incompressibility is a poorly known quantity experimentally, for the fact that some sort of theoretical modeling comes in these calculations. Apart from that, the other quantity with large uncertainty is the nucleon effective mass. The wide range of values determined from experiments of these two quantities motivates us to analyze and study the EOS with these variations. Further we also need to look into issues related to some indispensable elements, such as the $\sigma-$meson mass and the pion decay constant `$f_{\pi}$', while we want to achieve a proper framework for studying nuclear matter aspects. We recall that in the present work, our aim is to describe and correlate these physical quantities in a coherent and unified approach.

\section{Results and discussions}

The spontaneous breaking of chiral symmetry lends mass to the Hadrons and relates them to the vacuum expectation value (VEV) of the scalar field ($x_0$), which is what is shown in eqn. (2). Immediately, what follows from the third term in eqn. (2) is that, the VEV of the scalar field which has a minimum potential at $f_{\pi}$ is related to the vector coupling constant $C_{\omega}$ through the relation $x_0 = f_{\pi} = m_{\omega}/g_{\omega} = 1/ \sqrt{C_{\omega}}$. Thus the vector coupling constant is explicitly constrained from the vacuum value of the pion decay constant. Similarly the scalar meson mass can be given by  $m_{\sigma} = m~{\sqrt C_{\omega}/{\sqrt C_{\sigma}}}$. The model is then sternly constrained and exude the relationship between various quantities with the VEV of the scalar field. 

In the present calculation, we take the value of the saturation density to be $\rho_0 = 0.153 fm^{-3}$ \cite{compact}, which agrees with the observed charge and mass distribution of finite nuclei. Saturation density implies that the pressure of the system is zero and the system will remain in this state if left undisturbed. The binding energy per nucleon is fixed at an empirical value $B/A-m = -16.3$ MeV \cite{compact}. With the uncertainty in the nucleon effective mass at $\rho_0$, we calculate the parameters of the present model in the range $Y = m^{\star}/m = (0.75 - 0.90)$. In order to assure the existence of a lower bound for the energy, we demand that the coefficient `c' in the quartic scalar field term, remains positive. The corresponding related quantities, such as the pion decay constant, the scalar meson mass and the nuclear incompressibility are also calculated. The asymmetry energy coefficient is fixed at $J \approx 32$ MeV. The obtained parameters are enlisted in Table I, where the relationship between the vector coupling constant and the pion decay constant can be easily visualized. Stronger the vector coupling (repulsion), lower is the value of the chiral condensate and vice-versa. From the tabulated data, we find that the calculated sigma meson mass is predicted within ($340 - 700$) MeV for $m^{\star}/ m = (0.75 - 0.90)$. Although the values obtained from the analysis of neutron scattering off lead nuclei \cite{compact,nuclei} is consistent with the range $m^{\star}/m = (0.80 - 0.90)$, a lower nucleon effective mass is is known to reproduce the finite nuclei properties, such as the spin-orbit effects splitting correctly \cite{furn98}. Also, we find that as we move to higher effective mass region, the incompressibility of the matter starts to fall or the EOS gets softer. However, within the incompressibility range of K = (200 - 300) MeV, the present model predicts higher nucleon effective mass.

\begin{table*}
\caption{Parameter sets of the effective chiral model that satisfies the nuclear matter saturation properties such as binding energy per nucleon $B/A - m = -16.3 ~MeV$, nucleon effective mass $Y = m^{\star}/m = (0.75 - 0.90)$ and the asymmetry energy coefficient is $J \approx 32$ MeV at saturation density $\rho_0$ $=0.153 fm^{-3}$. The nucleon, the vector meson and the isovector vector meson masses are taken to be 939 MeV, 783 MeV and 770 MeV respectively and $c_{\sigma} = (g_{\sigma}/ m_{\sigma})^2$, $c_{\omega} = (g_{\omega}/ m_{\omega})^2$ and $c_{\rho} = (g_{\rho}/ m_{\rho})^2$ are the corresponding coupling constants. $B = b/m^2$ and $C = c/ m^4$ are the higher order constants in the scalar field. Other derived quantities such as the scalar meson mass `$m_{\sigma}$', the pion decay constant `$f_{\pi}$' and the nuclear matter incompressibility ($K$) at $\rho_0$ are also given.}
\begin{center}
\begin{tabular}{cccccccccc}
\hline
\hline
\multicolumn{1}{c}{set}&
\multicolumn{1}{c}{$c_{\sigma}$}&
\multicolumn{1}{c}{$c_{\omega}$} &
\multicolumn{1}{c}{$c_{\rho}$} &
\multicolumn{1}{c}{$B$} &
\multicolumn{1}{c}{$C$} &
\multicolumn{1}{c}{$m_{\sigma}$} &
\multicolumn{1}{c}{$Y$} &
\multicolumn{1}{c}{$f_{\pi}$} &
\multicolumn{1}{c}{$K$} \\
\multicolumn{1}{c}{ } &
\multicolumn{1}{c}{($fm^2$)} &
\multicolumn{1}{c}{($fm^2$)} &
\multicolumn{1}{c}{($fm^2$)} &
\multicolumn{1}{c}{($fm^2$)} &
\multicolumn{1}{c}{($fm^4$)}&
\multicolumn{1}{c}{(MeV)} &
\multicolumn{1}{c}{} &
\multicolumn{1}{c}{(MeV)} &
\multicolumn{1}{c}{($MeV$)} \\
\hline
1   &5.916  &3.207  &5.060  &1.411   &1.328    &691.379  &0.75  &110.185  &1098 \\
2   &6.047  &3.126  &5.087  &0.822   &0.022    &675.166  &0.76  &111.601  &916 \\
3   &6.086  &3.031  &5.107  &0.485   &0.174    &662.642  &0.77  &113.346  &809 \\
4   &6.005  &2.933  &5.131  &0.582   &2.650    &656.183  &0.78  &115.238  &737 \\
5   &6.172  &2.825  &5.155 &-0.261   &0.606    &635.287  &0.79  &117.403  &638 \\
6   &6.223  &2.709  &5.178 &-0.711   &0.748    &619.585  &0.80  &119.890  &560 \\
7   &6.325  &2.585  &5.200 &-1.381   &0.089    &600.270  &0.81  &122.740  &491 \\
8   &6.405  &2.451  &5.222 &-1.990   &0.030    &580.876  &0.82  &126.039  &440 \\
{\bf 9}   & {\bf 6.474}  & {\bf 2.323}  & {\bf 5.242} &{\bf -2.533}   &{\bf 0.300}    &{\bf 562.500}  &{\bf 0.83}  &{\bf 129.465}  &{\bf 391} \\
10  &6.598  &2.159  &5.265 &-3.340   &0.445    &536.838  &0.84  &134.378  &344 \\
{\bf 11}  &{\bf 6.772}  &{\bf 1.995}  &{\bf 5.285} &{\bf -4.274}   &{\bf 0.292}    &{\bf 509.644}  &{\bf 0.85}  &{\bf 139.710}  &{\bf 303} \\
12  &7.022  &1.823  &5.305 &-5.414   &0.039    &478.498  &0.86  &146.131  &265 \\
{\bf 13}  &{\bf 7.325}  &{\bf 1.642}  &{\bf 5.324} &{\bf -6.586}   &{\bf 0.571}    &{\bf 444.614}  &{\bf 0.87}  &{\bf 153.984}  &{\bf 231} \\ 
14  &7.865  &1.451  &5.343 &-8.315   &0.502    &403.303  &0.88  &163.824  &199 \\
15  &8.792  &1.249  &5.362 &-10.766  &0.354    &353.960  &0.89  &176.552  &168 \\
16  &7.942  &1.041  &5.388 &-6.908   &15.197   &339.910  &0.90  &193.437  &163 \\
\hline
\hline
\end{tabular}
\end{center}
\end{table*}

\begin{figure}[ht]
\begin{center}
\includegraphics[width=7cm,height=7cm,angle=0]{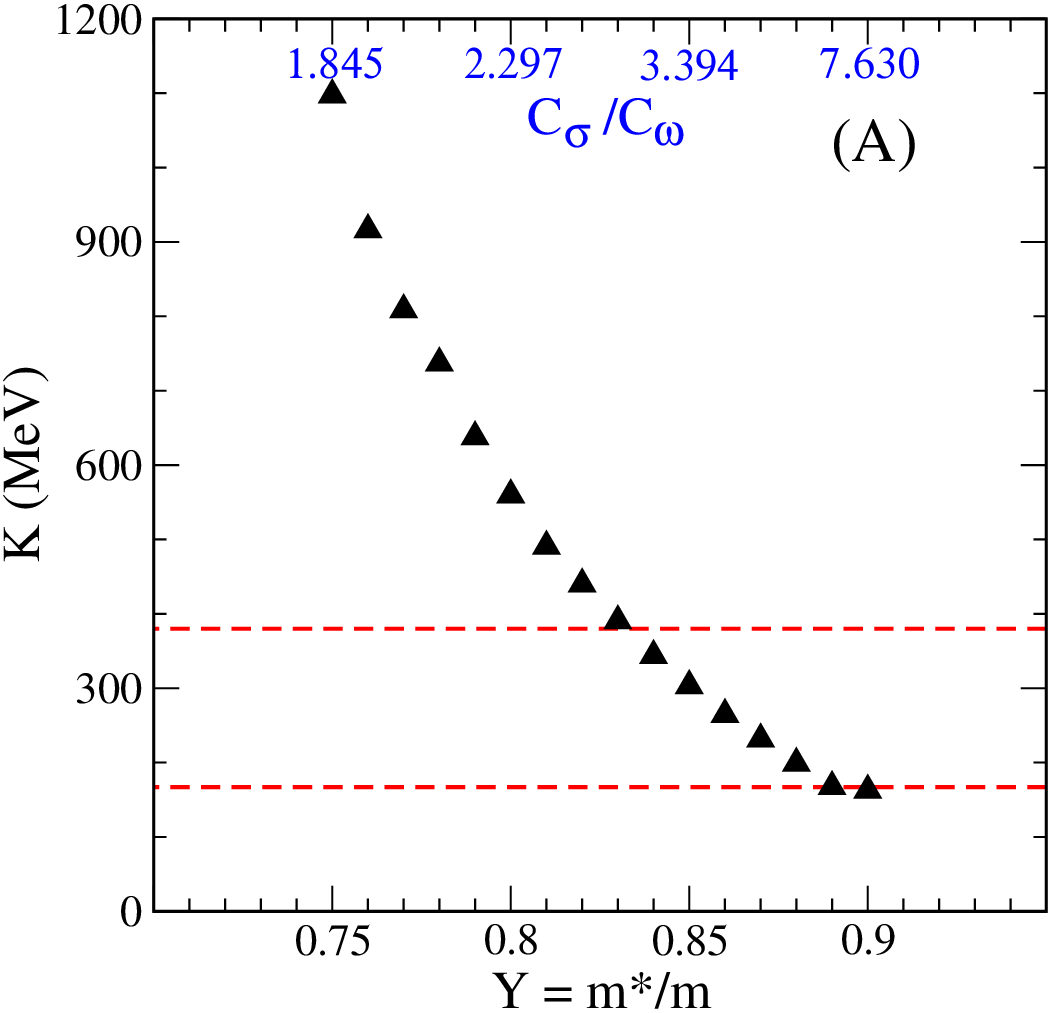}
\vskip 0.5in
\includegraphics[width=7cm,height=7cm,angle=0]{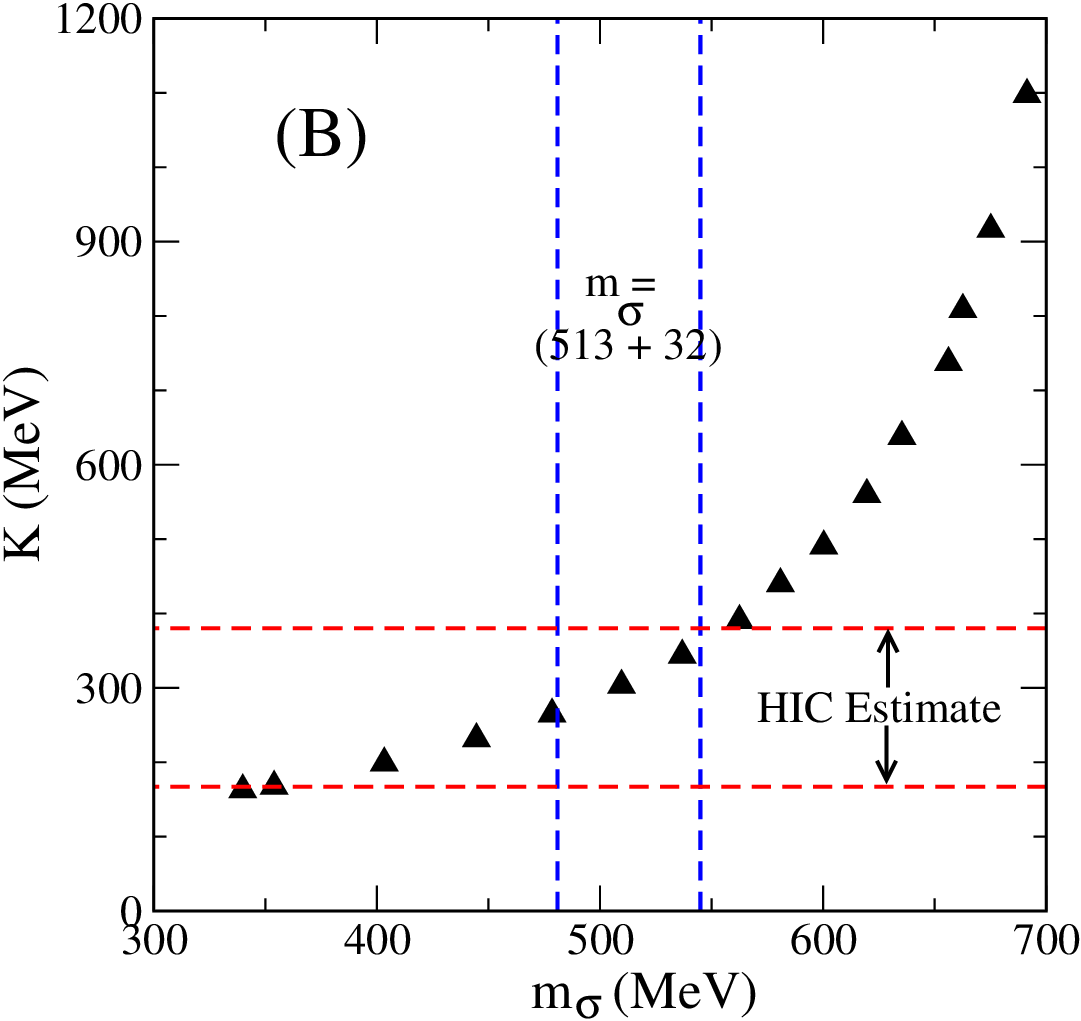}
\end{center}
\caption{(Color online)(A)- Nuclear matter incompressibility as a function of the nucleon effective mass for the parameters of the present model. Also plotted is the corresponding ratio of scalar to vector coupling on the opposite x-axis. (B)- Incompressibility as a function of obtained sigma meson mass for the parameter sets enlisted in Table I. The upper and the lower limit for incompressibility inferred from Heavy Ion Collision data \cite{data02} $K = (167-380)$ MeV is shown with horizontal red lines. Recent experimental scalar meson mass limit ($m_{\sigma} = 513 \pm 32$) MeV \cite{mura02} is depicted with blue vertical lines}
\label{effm}
\end{figure}

Nuclear matter saturation is a consequence of the interplay between the attractive (scalar) and the repulsive (vector) forces and hence the variation in the coupling strength effects other related properties as well. Fig. \ref{effm}(A) reflects the same, where we have plotted the nuclear incompressibility for the evaluated parameter sets of the present model as a function of the nucleon effective mass. For better correlation between them, the corresponding ratio of the scalar and vector coupling is also indicated. On comparison with the incompressibility bound inferred from heavy ion collision experiment (HIC)\cite{data02}, we find that the EOS with lower nucleon effective mass is ruled out. The present model favors EOS for which the nucleon effective mass $m^{\star}/m > 0.82$, i.e., the mass of the nucleon in the nuclear medium drops to less than $\approx 20 \%$ of its mass at $\rho_0$. Equivalently, the agreement with the experimental flow data in the density range $2 < \rho/\rho_0 < 4.6$ seem to favor repulsion (higher effective mass) in matter at high density. From the plot, it can be seen that the EOS becomes much softer with increasing ratio of $C_{\sigma}/C_{\omega}$.

Figure \ref{effm}(B) shows the variation of incompressibility as a function of scalar meson mass obtained for various parameter sets. Recent experimental estimate for scalar meson mass  $m_{\sigma} = 513 \pm 32$ MeV \cite{mura02} is compared with the present calculation. From the figure, we find that the EOS with $Y = (0.84 - 0.86)$ (Set 10, 11 \& 12; Table I) seems to agree with the combined constraint from the HIC flow data and the experimental meson mass range. Further, it is worth noticing that, lower the value of $m_{\sigma}$ lower is the value of incompressibility for matter and vice-versa. The heavy ion collision estimate seems to agree with $\sigma-$meson mass within ($350 - 550$) MeV. We know that the nuclear incompressibility and the $\sigma$-meson mass both are poorly determined quantities and therefore, some sort of correlation between the two will help to minimize the uncertainties around them.

\begin{figure}[ht]
\begin{center}
\includegraphics[width=7cm,height=7cm,angle=0]{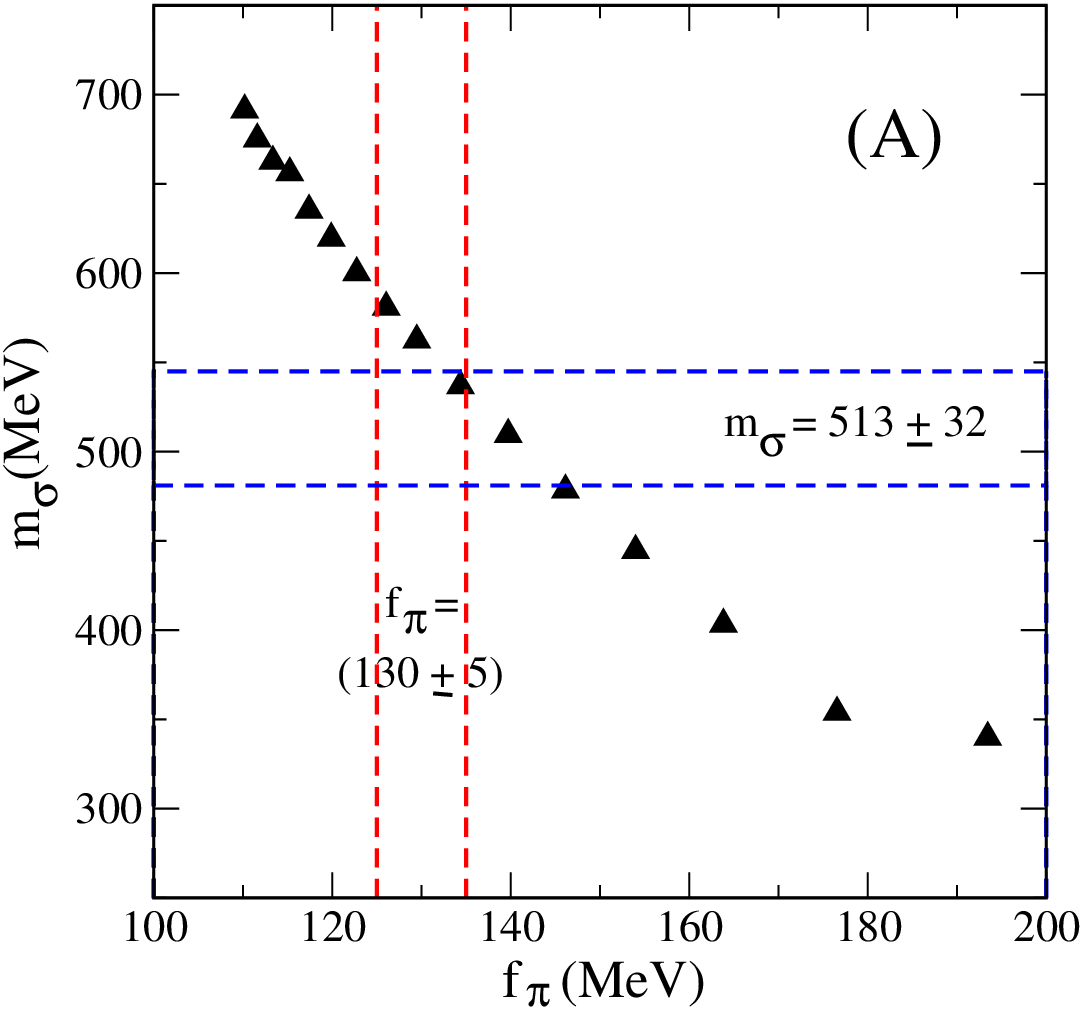}
\vskip 0.5in
\includegraphics[width=7cm,height=7cm,angle=0]{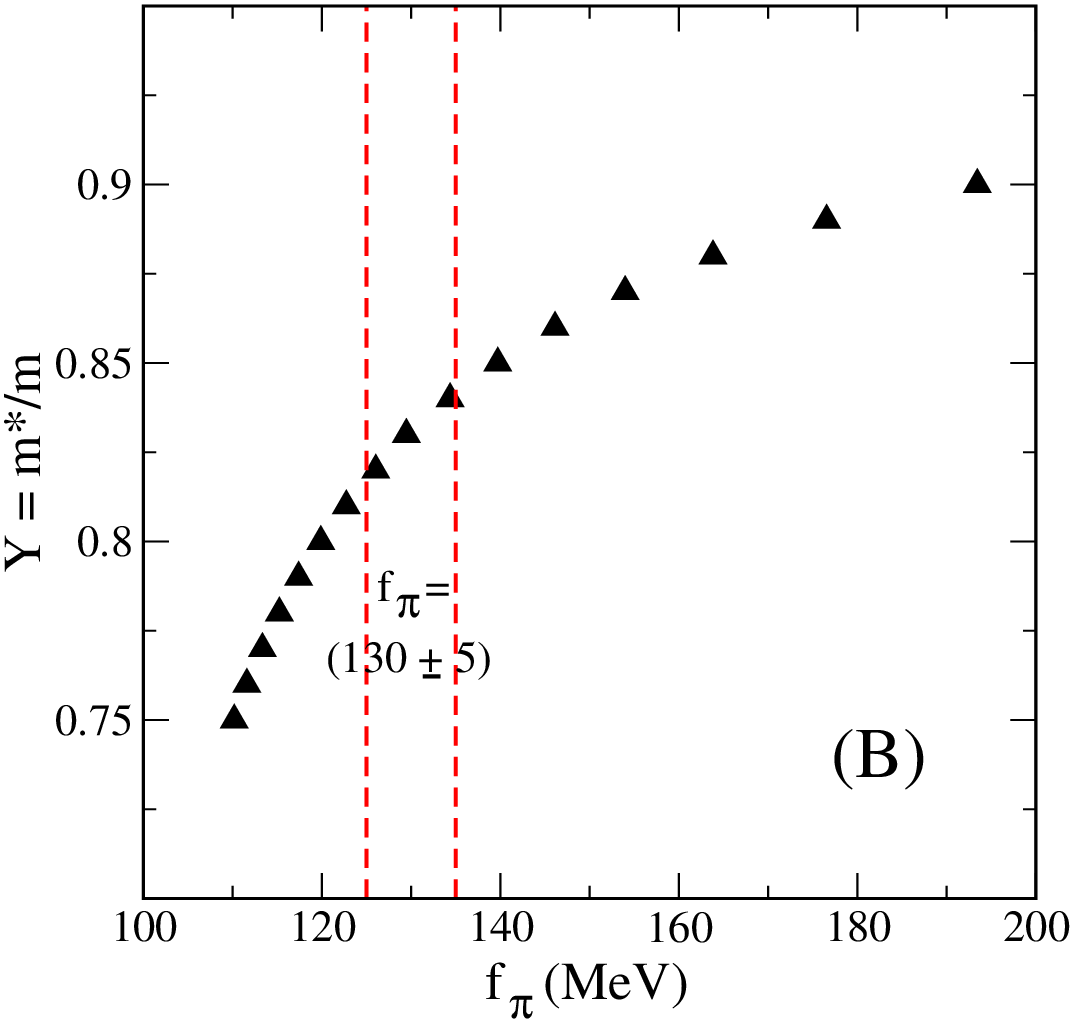}
\end{center}
\caption{(Color online)(A) - The scalar meson mass `$m_{\sigma}$' as a function of the vacuum value of the scalar field `$f_{\pi}$'. We take $f_{\pi} = 130 \pm 5$ MeV \cite{pdg}, which is shown with red vertical lines. The blue lines denotes the range of the mass of $m_{\sigma}$ from experiment \cite{mura02}. (B) - Ratio of nucleon effective mass to bare mass is plotted as a function of $f_{\pi}$.}
\label{fpi}
\end{figure}

Figure \ref{fpi}(A) shows the obtained sigma mass as a function of the vacuum value of the pion decay constant. It can be seen that both these physical quantities are inversely proportional to each other. A lower value of $f_{\pi}$ leads to a higher value of $m_{\sigma}$. However, the experimental bound of the pion decay constant seems to agree with slightly higher value of $m_{\sigma}$, which agree with the upper bound of the experimental bound on $m_{\sigma}$ \cite{mura02}. Precisely, the constraint of $f_{\pi}$ seems to agree with $m_{\sigma} \approx (560 \pm 22)$ MeV. Figure \ref{fpi}(B) shows the nucleon effective mass $Y = m^{\star}/m$ as a function of the pion decay constant. The constraint of $f_{\pi}$ agree with EOS with $Y = (0.82 -0.84)$ (Set 8, 9 \& 10; Table I). However the corresponding incompressibility falls in the range $K \approx (344 - 440)$ MeV, which is on the higher side of presently acceptable bounds (\cite{k2} - \cite{k6}). With combined constraints such as those on nuclear incompressibility inferred from HIC data \cite{data02}, the limits on sigma meson mass \cite{mura02}, the pion decay constant \cite{pdg} and the nucleon effective mass \cite{nuclei}, the best fit from the parameters can be extracted. Thus we choose parameter set 9, 11 and 13 of Table I to study the corresponding EOS and compare it with the experimental data \cite{data02} as well as other successful relativistic mean field models such as NL3 parameterization \cite{nl3} and the non-relativistic DBHF \cite{dbhf} calculations.

\section{Equation of state at $T = 0$}

\begin{figure}[ht]
\begin{center}
\includegraphics[width=7cm,height=7cm,angle=0]{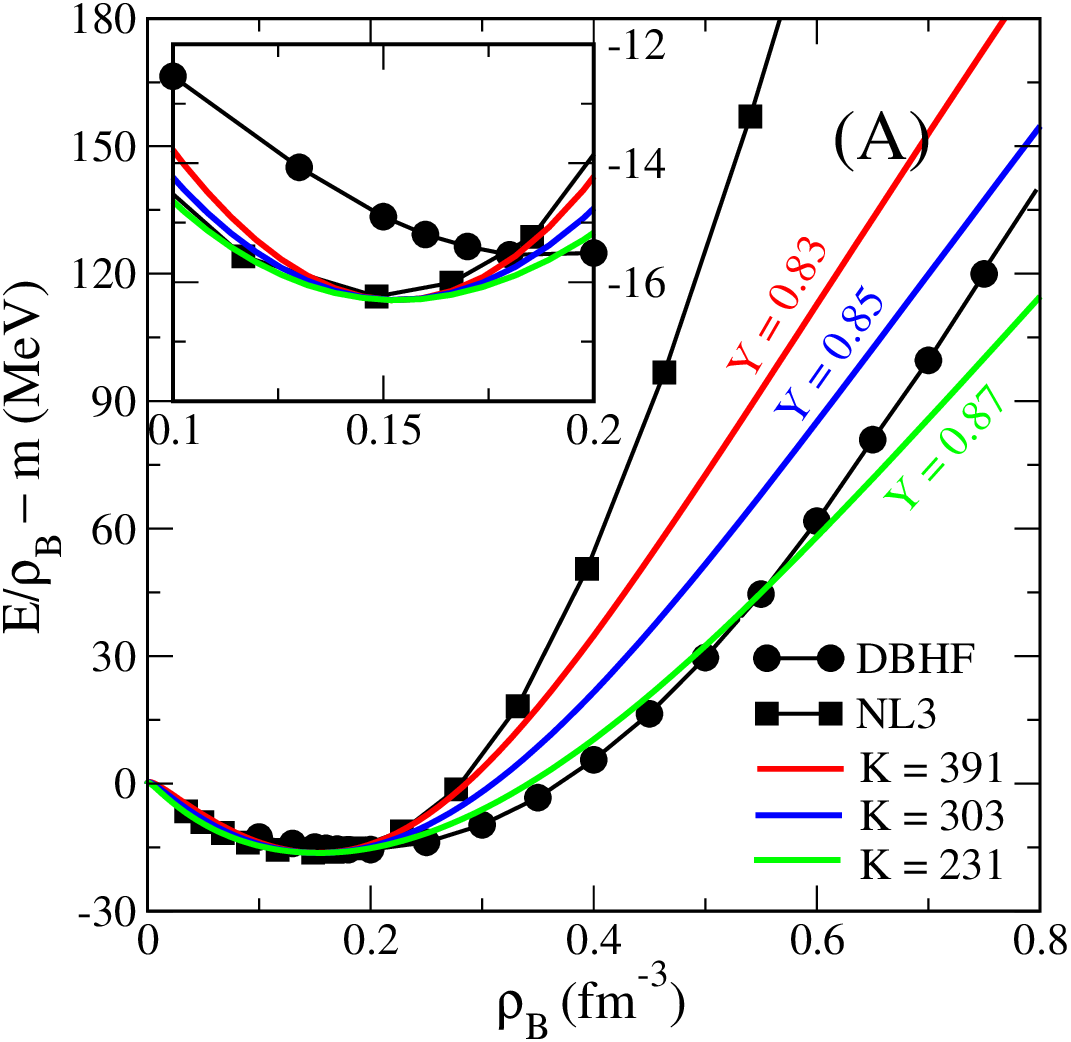}
\vskip 0.5in
\includegraphics[width=7cm,height=7cm,angle=0]{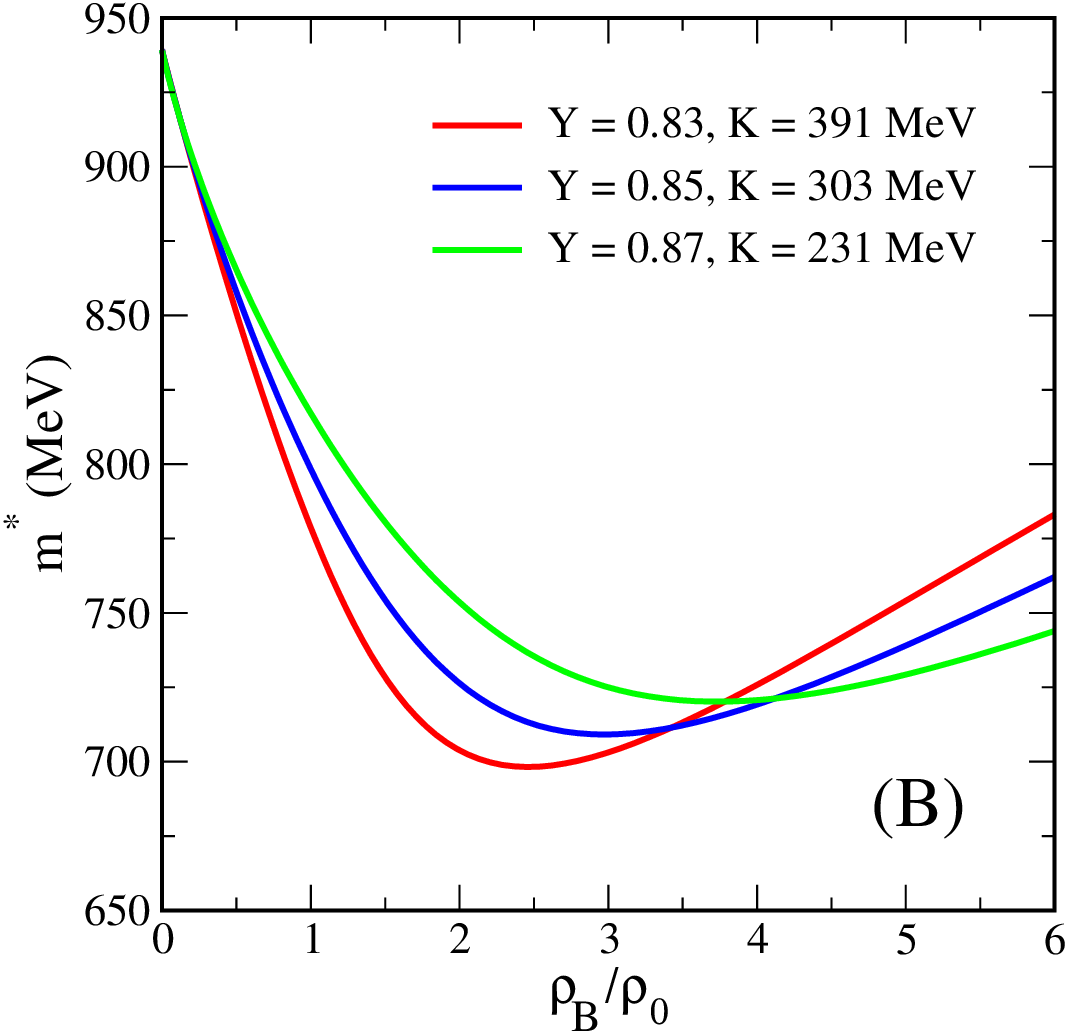}
\end{center}
\caption{(Color online) (A) - Binding energy per nucleon of symmetric nuclear matter plotted as a function of baryon density up to nearly $5 \rho_0$. For comparison, we also plot the same for NL3 parameter set from the Relativistic Mean Field theory \cite{nl3} as well as EOS from DBHF \cite{dbhf} in the non-relativistic domain. The inset plot displays the curve in the vicinity of nuclear saturation (B) - Variation of Nucleon effective mass in medium as a function of total baryon density for symmetric nuclear matter of the selected parameters of the present model.}
\label{snm}
\end{figure}

The selected parameters for further study is highlighted in bold fonts in Table I. The resulting energy per nucleon for symmetric nuclear matter is calculated for these parameters and is plotted in Fig. \ref{snm}(A). For comparison, we plot the same with NL3 parameterization from Relativistic Mean Field (RMF) calculations \cite{nl3} and also the non-relativistic realistic DBHF (Bonn-A) parameterization \cite{dbhf}. In the inset, the region of saturation density is magnified, where we find nice agreement within relativistic mean-field models near $\rho_0$. Although in case of NL3 parameter set, nuclear matter saturates at slightly lower density ($\rho_0 = 0.148 ~fm^{-3}$) than what we have taken in present calculation, the Binding energy per nucleon almost remains same ($\approx -16.3~ MeV$). In case of DBHF, nuclear matter saturates at still higher density. It is worth noticing that the incompressibility of the parameter chosen in the present model spans within (230 - 390) MeV, yet the resulting EOS seems to be soft at higher densities in comparison to that predicted by the NL3 parameter set which has $K = 271.6$ MeV. Incompressibility of nuclear matter is the measure of the degree of softness/ stiffness of the EOS. Conventionally, EOS with K $<$ 300 MeV are considered to be soft. But in the present case the EOS predicted by the effective model is relatively much softer than the NL3 parameterization although the value for the former is high enough. However this can be understood, if we look at the EOS in the vicinity of saturation density (Inset plot). The curve of NL3 seems to compare well with the EOS with $K = 231$ MeV below saturation density, but the energy predicted is much larger at higher densities. In contrast to that, the EOS predicted by the effective model gets softer at higher densities. It should be interesting to study the consequences of such behavior in the astrophysical context, especially on the global properties and structure of neutron stars at physically interesting densities (2 - 5 $\rho_0$) \cite{nkg,prak97}. 

In Fig. \ref{snm}(B), the nucleon effective mass in the nuclear medium is plotted as a function of baryon density up to 6$\rho_0$. This medium mass modification of nucleon in nuclear medium is a consequence of the Dirac field and forms an essential element for the success of the relativistic phenomenology. From the plot, it is interesting to see that the nucleon experiences repulsive forces in nuclear matter at higher densities ($\rho_B > 2 \rho_0$), as a result of which the nucleon effective mass increases again for the three cases that we study presently. A careful look into Table I reveals the relationship between the couplings (both scalar and vector) and the resulting nucleon effective mass. The model predicts a higher nucleon effective mass if the ratio of scalar to vector coupling is larger but the increase in $m^{\star}$ is slower thereafter, which reflects the dominance of attractive force at high densities. At saturation density, the present model results in much higher nucleon effective mass in comparison to the NL3 ($m^{\star}/m = 0.60$) and DBHF ($m^{\star}/m = 0.678$).

\begin{figure}[ht]
\begin{center}
\includegraphics[width=7cm,height=7cm,angle=0]{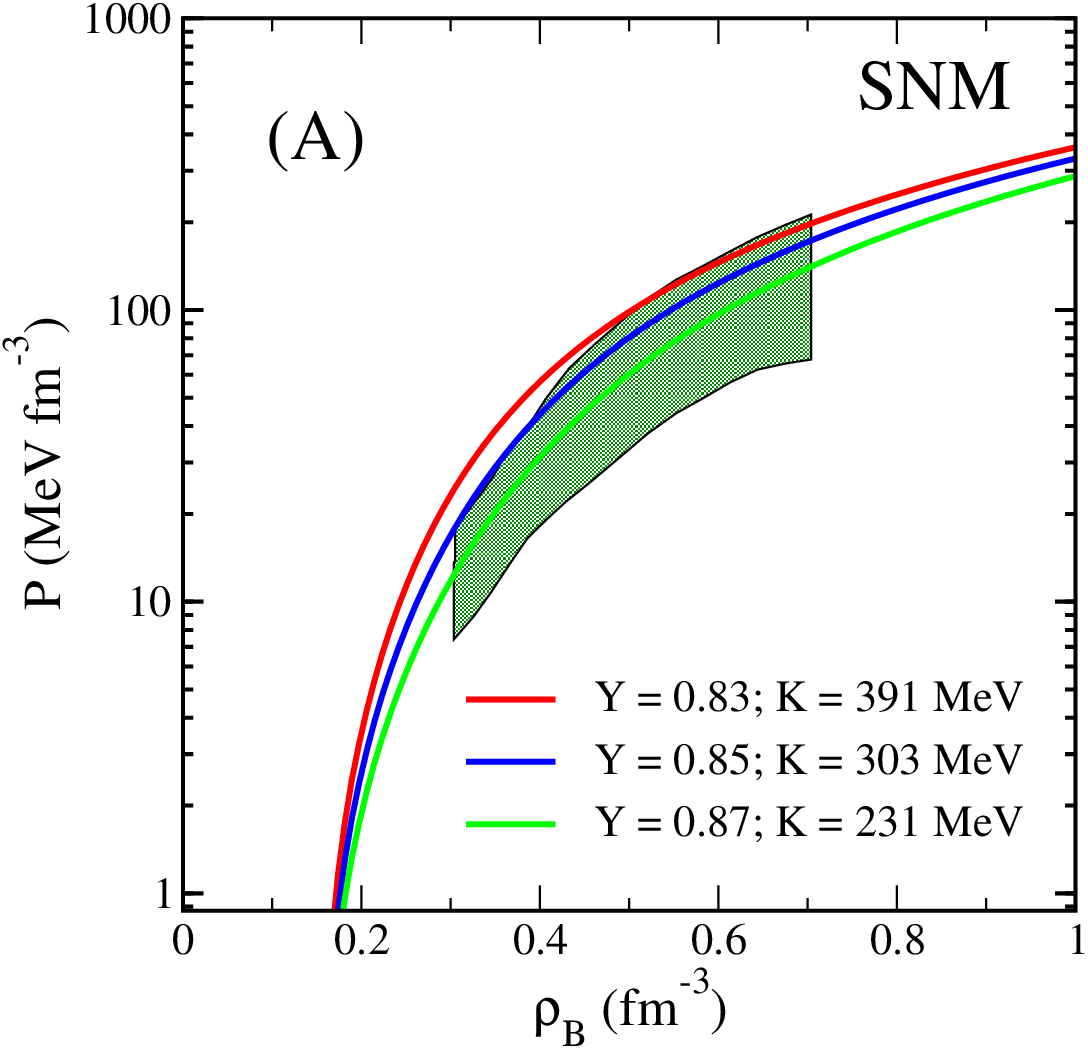}
\vskip 0.5in
\includegraphics[width=7cm,height=7cm,angle=0]{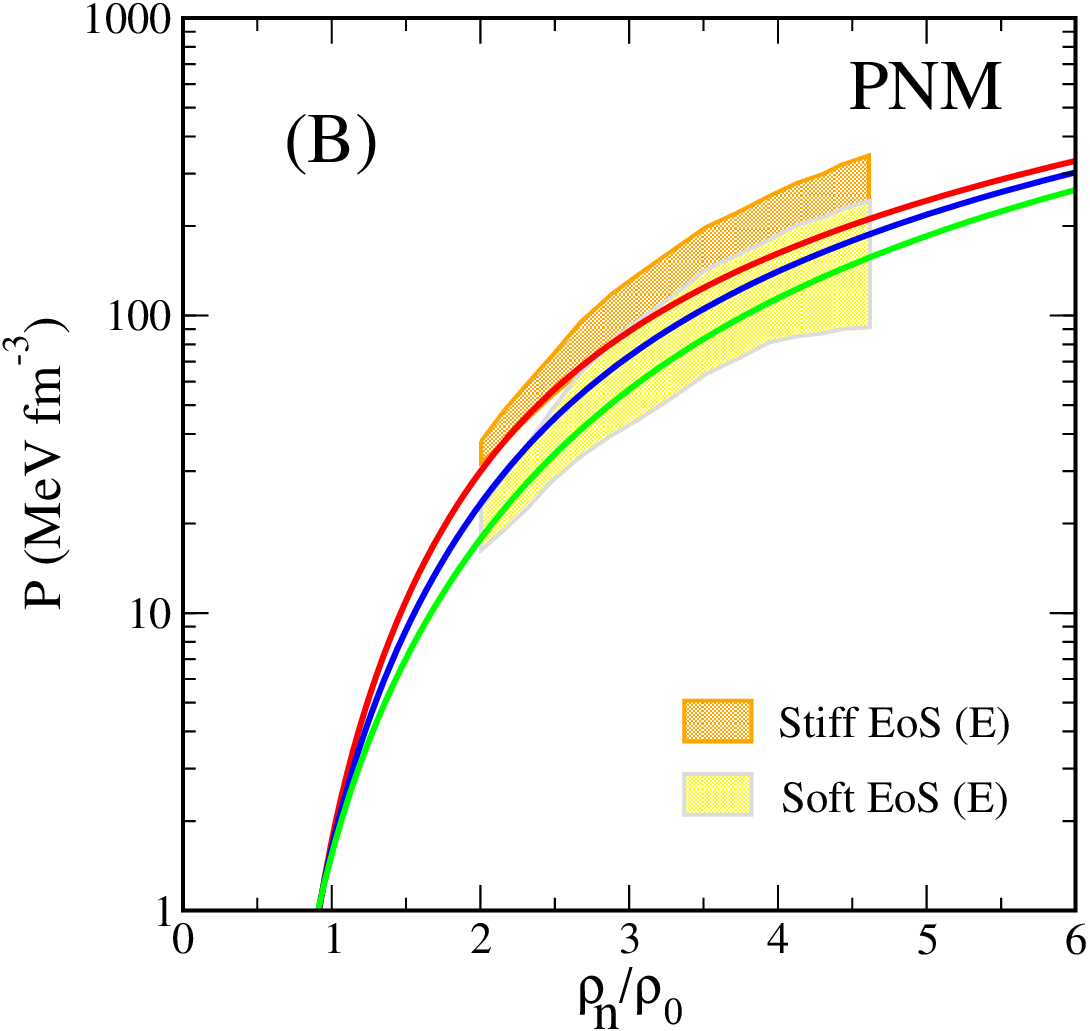}
\end{center}
\caption{(Color online) Comparison of the Heavy Ion Collision estimate \cite{data02} with the theoretical prediction of the effective model, (A) for Symmetric nuclear matter (SNM) case and (B) for Pure Neutron matter (PNM) case.}
\label{expt}
\end{figure}

Fig. \ref{expt}(A) displays the pressure as a function of baryon density up to nearly 6$\rho_0$ for the selected parameters of the model for symmetric nuclear matter. The shaded region corresponds to the experimental HIC data \cite{data02} for symmetric nuclear matter (SNM). Among the three theoretical calculations shown, the EOS with Y = 0.85 \& 0.87 agree very well with the collision data. Precisely, the third set (K = 231 MeV) completely agree with the flow data in the entire density span of $2 < \rho_B/\rho_0 < 4.6$. We now proceed to calculate the EOS of Pure Neutron matter (PNM) by taking the spin degeneracy $\gamma = 2$ in eqn. (8) and (9). The inclusion of $\rho-$meson doesn't seem to affect the EOS substantially and so we refrained from that. In Fig. \ref{expt}(B), the case of pure neutron matter (PNM) is compared with the experimental flow data. The experimental flow data is categorized in terms of stiff or soft based on whether the density dependence of the symmetry energy term is strong or week \cite{pra88}. The EOS predicted by the present model seems to rather lie on the softer regime. However, the EOS with $Y = 0.87, K = 231$ MeV though satisfy the combined constraint rather well, is not consistent with the vacuum value of the pion decay constant.

\section{Summary and conclusions}

The effective chiral model provides a natural framework to interlink the standard state properties of nuclear matter with the inherent fundamental constants such as the pion decay constant and the $\sigma-$meson mass within a unified approach. With this motivation, the parameters of the model are evaluated in the mean-field ansatz by fixing the nuclear matter saturation properties defined at $\rho_0$ and varying the nucleon effective mass in the range $Y = m^{\star}/m = (0.75 - 0.90)$. Thus the resulting equation of state not only satisfies the saturation properties reasonably well but also relates the various aforesaid fundamental quantities with that of the vacuum value of the scalar field constant. One of the unique features of the model is that the mass of the vector meson ($m_{\omega}$) is generated dynamically, as result of which the effective mass of the nucleon acquires a density dependence on both the scalar and vector fields. The interplay between this scalar and vector forces results in the increase of $m^{\star}$ at $\approx 3 \rho_0$ (Fig. 3(B)), as a consequence of which the resulting EOS is much softer at higher densities.

We also discussed the implication of imposing fundamental constraint on the evaluated model parameters. Among various derived quantities, we find that the pion decay constant is experimentally well known quantity in comparison to the nuclear incompressibility and $\sigma-$meson mass, that can put stringent constraint on the model parameters. Employing this constraint ($f_{\pi} = 130 \pm 5$ MeV) rather leave us with few options among the wide range of parameters enlisted in Table I, while the observed range of $\sigma-$meson mass do not rule out any. Experimentally determined effective mass from scattering of neutron over $Pb$ nuclei \cite{nuclei} seems to go well with the present model. Both of them favor higher value for nucleon effective mass. In the present calculation we find that a higher nucleon effective mass is endowed with reasonable incompressibility too. However, the parameter that agree well ($m^{\star} = 0.83 m$; Set 9) with the limits of $f_{\pi}$ has incompressibility in the upper bound of the value inferred from the flow data. On a comparative analysis of the resulting EOS with that of the HIC data for symmetric nuclear matter as well as pure neutron matter, parameter set with $Y = 0.85; K \approx 300$ MeV seems to be the ideal parameterization of the present model. The resulting scalar meson mass $m_{\sigma} \approx 510 MeV$, is also consistent with the experimentally observed masses \cite{mura02,aitala01,ishida01}. Further, a higher value of incompressibility $K \approx 300 $ MeV is known to predict correctly the isoscalar giant resonance energies in medium and heavy nuclei in the relativistic framework \cite{ma97}. On account of the aforesaid arguments and constraints, the model seems to work very well within present approach. However, the predictability of the model needs to be tested at finite temperature and high densities. Work is in progress in this direction \cite{tkj09}. In this regard, it will also be interesting to study the medium effects on the underlying couplings \cite{rmfden99} as well as the on the meson masses \cite{meson} and the pion decay constant \cite{pion}.

\section{Aknowledgment}

One of the authors HM would like to thank Institut for Theoretische Physik, University of Frankfurt for warm hospitality and Alexander von Humbolt foundation, Germany for support during this period.

\end{document}